\documentclass[12pt]{article}
\usepackage{amsmath}
\usepackage{graphicx,psfrag,epsf}
\usepackage{enumerate}
\usepackage{natbib}
\usepackage{textcomp}
\usepackage[hyphens]{url} 
\usepackage{hyperref}

\newcommand{\blind}{0}

\usepackage{color}
\usepackage{fancyvrb}

\DefineVerbatimEnvironment{Highlighting}{Verbatim}{commandchars=\\\{\}}
\usepackage{framed}
\definecolor{shadecolor}{RGB}{248,248,248}
\newenvironment{Shaded}{\begin{snugshade}}{\end{snugshade}}

\newcommand{\AttributeTok}[1]{\textcolor[rgb]{0.77,0.63,0.00}{#1}}

\newcommand{\DecValTok}[1]{\textcolor[rgb]{0.00,0.00,0.81}{#1}}

\newcommand{\FunctionTok}[1]{\textcolor[rgb]{0.00,0.00,0.00}{#1}}

\newcommand{\NormalTok}[1]{#1}

\newcommand{\OtherTok}[1]{\textcolor[rgb]{0.56,0.35,0.01}{#1}}

\newcommand{\SpecialCharTok}[1]{\textcolor[rgb]{0.00,0.00,0.00}{#1}}

\newcommand{\StringTok}[1]{\textcolor[rgb]{0.31,0.60,0.02}{#1}}

\providecommand{\tightlist}{%
  \setlength{\itemsep}{0pt}\setlength{\parskip}{0pt}}

\usepackage{longtable,booktabs,array}
\usepackage{calc} 
\usepackage{etoolbox}
\makeatletter
\patchcmd\longtable{\par}{\if@noskipsec\mbox{}\fi\par}{}{}
\makeatother
\IfFileExists{footnotehyper.sty}{\usepackage{footnotehyper}}{\usepackage{footnote}}
\makesavenoteenv{longtable}

\newcommand{\mdids}[1]{https://www.introdata.science}
\newcommand{\mcrids}[1]{https://introds-2020.netlify.app}
\newcommand{\dsbox}[1]{https://datasciencebox.org}
\newcommand{\mdidsgithub}[1]{https://github.com/ics80-fa21/website}
\newcommand{\mcridsgithub}[1]{https://github.com/ids-s1-20}

\begin{document}

\def\spacingset#1{\renewcommand{\baselinestretch}%
{#1}\small\normalsize} \spacingset{1}


\if0\blind
{
  \title{\bf Tools and Recommendations for Reproducible Teaching}

  \author{
        Mine Dogucu \footnote{\href{mailto:mdogucu@uci.edu}{\nolinkurl{mdogucu@uci.edu}}} \\
    Department of Statistics, University of California Irvine\\
     and \\     Mine \c{C}etinkaya-Rundel \footnote{\href{mailto:mc301@duke.edu}{\nolinkurl{mc301@duke.edu}}} \\
    Department of Statistical Science, Duke University and RStudio\\
      }
  \maketitle
} \fi

\if1\blind
{
  \bigskip
  \bigskip
  \bigskip
  \begin{center}
    {\LARGE\bf Tools and Recommendations for Reproducible Teaching}
  \end{center}
  \medskip
} \fi

\bigskip
\begin{abstract}
It is recommended that teacher-scholars of data science adopt reproducible workflows in their research as scholars and teach reproducible workflows to their students. In this paper, we propose a third dimension to reproducibility practices and recommend that regardless of whether they teach reproducibility in their courses or not, data science instructors adopt reproducible workflows for their own teaching. We consider computational reproducibility, documentation, and openness as three pillars of reproducible teaching framework. We share tools, examples, and recommendations for the three pillars.
\end{abstract}

\noindent%
\vfill

\newpage
\spacingset{1.45} 

\hypertarget{introduction}{%
\section{Introduction}\label{introduction}}

For teacher-scholars of data science, the literature suggests two distinct areas of reproducibility.
First, as scholars, the literature extensively focuses on the adoption of reproducible workflows in research including the American Statistical Associations' reproducible research recommendations \citep{ASAreproducibility}.
Second, for teachers, the literature focuses on teaching reproducible practices including literate programming \citep{baumer2014, dvorak2019}, documentation \citep{ball2012}, version control \citep{fiksel2019, beckman2021}, and open science practices \citep{toelch2018}.
In this paper, we propose a third dimension to reproducibility practices for teacher-scholars and consider reproducible teaching practices.
Our vision for the modern teacher-scholars is having consistent reproducibility practices in how they conduct research, what they teach to students, and how they prepare teaching materials.
We distinguish the three aspects as reproducible research, teaching reproducibility, and reproducible teaching.
In this paper, we mainly focus on the last point.

To distinguish the three dimensions of teacher-scholars' practices, consider a hypothetical teacher-scholar who uses R for their research, teaches students how to use a graphing calculator for calculating statistics, and uses a slideshow software to prepare their teaching materials.
In this hypothetical scenario, three are separate
tools and workflows for research (R), for learning (calculator), and for teaching (slideshow software).
Previously statistical software have been classified in two categories: tools for learning and doing statistics \citep{biehler1997, mcnamara2016}.
In this manuscript we consider a third category: tools for teaching.
We share tools for teaching from a reproducibility point of view and share practices that can provide consistency between doing, teaching, and learning statistics and data science.

We suspect that data science courses include greater number of files and folders than a traditional statistics courses since data science courses focus more on computing.
We also suspect decreased reliance on paper-based assignments (e.g.~problem sets) with recent remote learning experiences which could have potentially also increased the number of digital files that instructors manage.
In addition, with the rise of interest in data science courses, the courses are getting larger thus instructors need to manage an increasing number of students' files as well.
Recommendations for good computing practices for file and data management and documentation often focus on research \citep{wilson2017}.
However, some of these best practices can also be extended to teaching.

Role modeling and teaching by example are effective teaching methods \citep{cruess2008, crowe2004, howard2021}.
In a typical data science classroom, even if students are taught reproducibility practices, they are often not exposed to the professor's research practices where they could possibly learn by observing the professor's workflow.
The lack of exposure to the professor's research workflow deprives students of a role model in reproducibility practices.
Adopting a reproducible workflow in the preparation and delivery of teaching materials and artifacts can set an example to students and provide them with further exposure to the tools they use for their own learning.
Setting an example to students in reproducibility practices can be possible if the set or the subset of tools that are used for teaching reproducibility are also used for reproducible teaching.

Teaching materials are often shared with others, even though the nature of sharing is generally different than that of research outcomes.
For instance, a course instructor might share teaching materials with their teaching assistants and graders.
In addition, instructors, especially those early in their careers, might inherit course materials from a colleague.
Similarly, many instructors make their teaching materials available online.
Some examples include D'Agostino McGowan's \emph{Statistical Learning} course \citeyearpar{mcgowancourse}, Baumer's \emph{Introduction to Data Science} course \citeyearpar{baumercourse}, and Blitzstein's \emph{Probability} course\citeyearpar{blitzsteincourse}.
In all teaching situations where teaching materials are shared, reproducibility becomes important in order for other instructors to be able to utilize the shared resources in their own teaching.
Reproducible teaching resources shared with the right permissions can also help advance statistics and data science education by allowing instructors to use, re-use, improve each others' resources.
As more reproducible teaching materials are shared more is learned from each other about teaching.

In the following sections, we define what reproducible teaching is by establishing a framework and then we provide a set of tools and recommendations for statistics and data science educators to achieve reproducibility in their teaching.
The framework we present is agnostic of programming language taught.
Our tooling examples are centered around R \citep{R}, however, these tools can be used to support teaching other computing languages such as Python, Julia.
SQL, etc.

\hypertarget{framework-for-reproducible-teaching}{%
\section{Framework for reproducible teaching}\label{framework-for-reproducible-teaching}}

We propose a framework for reproducible teaching around the following three pillars:

All teaching materials should be

\begin{itemize}
\tightlist
\item
  computationally reproducible,
\item
  well-documented, and
\item
  open.
\end{itemize}

\hypertarget{computational-reproducibility}{%
\subsection{Computational reproducibility}\label{computational-reproducibility}}

In research, computational reproducibility is defined as having information about code, data, and software to replicate the findings which usually is beyond what is provided in a traditional research article \citep{stodden2013b}.
We define computationally reproducible teaching materials as materials that can be reproduced identically at a later time by the same instructor (e.g., following semester) or by other instructors (i.e., on a different machine with a different setup and at any point in time).
The recommendations for achieving computational reproducibility for teaching materials are very similar to the recommendations for achieving this goal for research materials:

\begin{itemize}
\tightlist
\item
  \textbf{Literate programming:} Prefer tooling that uses plain text for prose, integrates code into the same document, and produces output that brings the two pieces together with literate programming.
\item
  \textbf{Raw data:} Keep data in the raw form you find or collect it, and record any steps to process it to prepare it for teaching.
\item
  \textbf{File organization:} Keep all files needed to reproduce the materials in a single folder and use a naming convention that is both machine- and human-readable.
\item
  \textbf{Version control:} Keep all files under version control.
\end{itemize}

\hypertarget{documentation}{%
\subsection{Documentation}\label{documentation}}

\citet{stodden2013} lists ``well-documented code'' as one of the requirements for research to be considered reproducible.
The same standard can be applied to teaching materials as well.
In a repository of teaching materials, the data, each of the folders, and software required to reproduce the materials should all be documented and the documentation, along with the code, should follow a particular style guide.

\begin{itemize}
\tightlist
\item
  \textbf{Data documentation:} The primary method of data documentation is a codebook, which should be written in plain text.
\item
  \textbf{File documentation:} The course folder as well as each of its top-level sub-folders should have a README file that explains what is included in that folder and outlines any steps to reproduce the contents.
\item
  \textbf{Software documentation:} Source code alone is rarely sufficient for reproducibility. For teaching materials to be reproducible one needs to also know the software (including versions) required to reproduce the results.
\item
  \textbf{Style guide:} A style guide is a set of guidelines that standardizes the formatting of writing. Even though a style guide is not a requirement for reproducibility, complying to a style guide helps with reading code, documentation, and overall organization easier.
\end{itemize}

\hypertarget{openness}{%
\subsection{Openness}\label{openness}}

Many reproducible research guidelines recommend open sharing of materials \citep{ibanez}, and so do we as part of our reproducible teaching framework.
For full reproducibility by others, all source code should be openly shared, however various instructors might be comfortable with various levels of openness ranging from full open source to just open access.
Regardless of the level, important considerations are the license with which the materials are shared and how and where they're hosted.

\begin{itemize}
\tightlist
\item
  \textbf{Licensing:} It is essential that all work shared openly is released with a license that makes it clear how others can reproduce and reuse the material. The choice of the license depends on a variety of factors, such as whether you want to allow commercial use of the materials or limit their use to noncommercial educational settings, how derivatives can be shared, etc. Additionally, licenses that are most appropriate for course materials are different than those for code (e.g., an R package that supplements the teaching materials.)
\item
  \textbf{Hosting:} The course materials, if shared openly with the general public would need to be hosted on a website. Depending on the tooling that the instructors use different hosts can be utilized.
\end{itemize}

\hypertarget{tools-and-recommendations-for-reproducible-teaching}{%
\section{Tools and recommendations for reproducible teaching}\label{tools-and-recommendations-for-reproducible-teaching}}

\hypertarget{computational-reproducibility-1}{%
\subsection{Computational reproducibility}\label{computational-reproducibility-1}}

\hypertarget{literate-programming}{%
\subsubsection{Literate programming}\label{literate-programming}}

A typical workflow for teaching data science includes, at the very minimum, some data analysis and presenting the data analysis process to learners.
This process may involve working across multiple file formats such as copying-pasting from statistical software (e.g., R, Python) to a text editor (e.g., Microsoft Word, PowerPoint, LaTeX).
However, this workflow is error-prone and is likely to result in copy-paste mistakes.
In addition, the results that are copy-pasted can be difficult to reproduce since the code that produced them may not be straightforward to track down.
Literate programming tools like R Markdown \citep{rmarkdown} and Jupyter notebooks \citep{jupyter} provide an opportunity to bring together analysis and writing components in a single file.

Preparing teaching materials using R Markdown enables reducing the number of files necessary to maintain.
For instance, instructors do not need to provide notes and R code separately but can unify the two in an R Markdown document.
R Markdown also extends to a wider range of packages that can easily be adopted in the classroom.
For instance, the \textbf{xaringan} package \citep{xaringan} can be utilized by instructors for making presentation slides. In addition, for preparing course materials or writing lecture notes as books, instructors can use the \textbf{bookdown} package \citep{bookdown}. For developing course websites or students' data science portfolios, as well as blogs and scientific writing on the web, the \textbf{distill} \citep{distill} and \textbf{blogdown} \citep{blogdown} packages provide easy to use templates.

Overall, R Markdown and its extension packages allow for writing reports, course materials, presentations in a reproducible way that minimizes human error related to copying code and/or text.
Using R Markdown for teaching material preparation does not necessitate teaching R Markdown.
The aforementioned packages can also be used by instructors who teach courses in a language other than R, courses that are software-agnostic, or those that utilize point-and-click online applets.

Instructors who already teach R Markdown to students may also choose to share their R Markdown files for teaching materials openly with the students even if they do not share with the broader public.
This allows the students to see that the tool they are learning is in fact useful.
When instructors share their R Markdown files, students get an opportunity to peek into more advanced features, if they choose to.
For instance, due to the hectic curriculum of the introductory data science classes, we do not focus on teaching styling R Markdown files but rather rely on built-in styles.
From time to time, we receive questions from students about how we could do a certain thing on our slides.
Having the source code of the R Markdown file openly accessible means that we can easily refer them to the source code.
There have been a few occasions when we received assignment submissions where students had taken their R Markdown skills above and beyond what is covered in our courses.

\hypertarget{raw-data}{%
\subsubsection{Raw data}\label{raw-data}}

Teaching statistics and data science generally requires lots of data -- small data, big data, toy data, simulated data, and anything in between.
This means that data management best practices are just as applicable to the creation and maintenance of teaching materials as they are to research.

Often times the data analysis steps featured in a single lecture might require the data to be pre-processed to make a particular point.
We recommend keeping the raw data file and scripting the pre-processing steps so that they can be easily reconstructed from the raw data.
If the lecture slides are created with literate programming (e.g., using the R Markdown based xaringan package) the pre-processing steps that are not to be shown to students within the context of that lecture can be put in a code chunk with \texttt{include\ =\ FALSE} as a chunk option.
This option means the code will run, but neither the code itself nor the results, or any warnings, messages, etc. from it will not be shown to students.

Sometimes the pre-processing needed to get data in shape to be featured in a lecture or assignment can be computationally intensive.
For example, maybe you need to scrape the data first from thousand of webpages or query it out of a database and reshape it.
Code required to accomplish these tasks would not be ideal to include in a code chunk in the R Markdown document that serves as the source code for slides, assignment sheets, etc.
In these cases, we recommend creating a separate \texttt{data} folder as a sub-directory of the lesson or assignment being prepared and include all of the code required to fetch the data or transform and clean the raw data in a script file.
This approach will not only make it easier to update the data at a later time but will also make it easier to fix any errors that might have been introduced during the data pre-processing stage.

The suggestions above are for working with real data, which is the type of data everyone always recommends for teaching.
But we also all know that sometimes you need a toy dataset or simulated data to make a particular point in a short amount of time, and sometimes in a small space on a single slide, when teaching.
If you're creating a toy dataset of just a few rows, we recommend using a function like \texttt{tibble::tibble()} or \texttt{tibble::tribble()} to create the dataset in a code chunk.
If you're simulating data or randomly sampling, you can also do this in a code chunk, but we recommend setting a seed so that the same simulated data can be achieved regardless of how many times the document is compiled.

\hypertarget{file-organization}{%
\subsubsection{File organization}\label{file-organization}}

Working with data and programming requires a basic understanding of working directories.
Working directories, how files are managed within the directories, and file paths become more important when multiple people work on the same set of files.
In a classroom setting, this happens quite often as, at the very least, a file is shared with students, teaching assistants or other instructors.
For instance, if a document provided by the instructor contains a statement such as \texttt{read.csv("C:\textbackslash{}Users\textbackslash{}instructor-username\textbackslash{}Documents\textbackslash{}somedata.csv")}, others using this file would have to change the file path provided in the code to match where data lives locally on their own computer.
Sharing code with file paths that could work on anyone's computer without making changes to the code is possible.
Consistency of file paths across different users are often achieved in software through utilization of projects.
Once a project-based workflow is adopted, then relative file paths can be utilized.
RStudio supports project-based workflows.

\begin{figure}

{\centering \includegraphics[height=0.4\textheight]{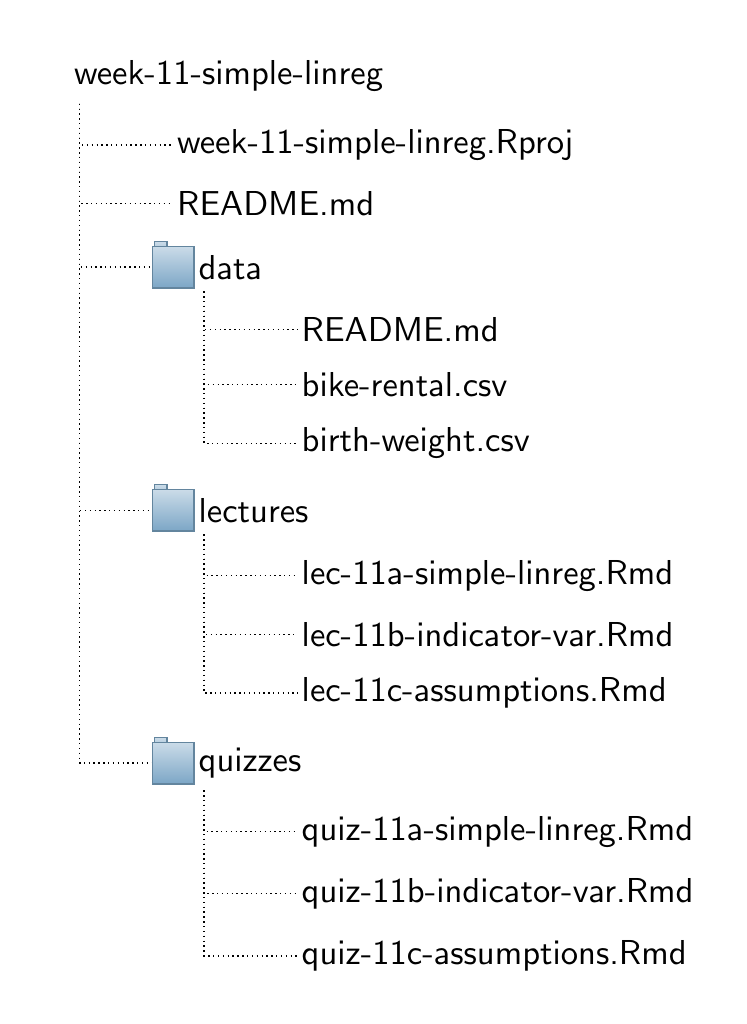} 

}

\caption{A folder template for weekly course materials}\label{fig:directory}
\end{figure}

We explain project-based workflow with an example.
In Figure \ref{fig:directory}, we present a template folder that mimics a typical weekly folder in one of our classes.
We release course materials weekly to students.
In each week, we release the necessary data files, lecture (starter) code, and quiz questions that students work on.
In this figure, we have a project file \texttt{week-11-simple-linreg.Rproj}.
We always use this project file to launch the project in the RStudio IDE and then access files within this folder via the Files pane in the IDE.
Launching the project in this manner starts an R session in which the working directory is set to the project folder and all users who have this project folder can use \texttt{read.csv("data/bike-rental.csv")} to read in data without having to change file paths.

Depending on the complexity of the project and the hierarchy of files in the project folder, one might need to use multiple ways to define the path to a certain file.
This is especially true when R Markdown is part of the toolkit, since when knitted the working directory of the R Markdown file is the directory where the file lives while when the code from the R Markdown document is ran interactively, the working directory is the project directory.
The \texttt{here()} function from the \textbf{here} package \citep{here} is especially useful for addressing this wrinkle.
It essentially sets the file path to the root of the project directory no matter where the code is being ran from (as part of the Knit process or interactively in the Console) For instance, the following code will read the \texttt{bike-rental.csv} file in an R script and in an R Markdown file as it sets the file path to the \texttt{week-11-simple-linreg} folder.

\begin{Shaded}
\begin{Highlighting}[]
\FunctionTok{read.csv}\NormalTok{(here}\SpecialCharTok{::}\FunctionTok{here}\NormalTok{(}\StringTok{"data"}\NormalTok{, }\StringTok{"bike{-}rental.csv"}\NormalTok{))}
\end{Highlighting}
\end{Shaded}

For naming files, \citet{bryantalk} recommends three principles: that the file names are machine-readable, human-readable, and play well with default ordering.
Based on these principles, for teaching, a natural enumeration of files can be based on the timing in the academic calendar.
In other words, file names can reveal the timing in the calendar (e.g.~week of the academic term).
In addition, files names can reveal the type of activity and/or assessment, and the topic of the file content.
For instance, \texttt{quiz-01a-intro-data.Rmd} reveals that this document is a R Markdown file for a quiz in the first week (01), in the first class (a) on the topic introduction to data.
Similarly \texttt{lec-11c-log-regression.Rmd} reveals that this document is a R Markdown file for lecture in the eleventh (11) week, in the third class (c) on the topic logistic regression.
These file names are both machine- and human-readable.
Instructors may be tempted to abbreviate the days of the week M, T, W, R, F to be used in naming course material files, however, these letters do not play well with default ordering.
Thus, alphabetically ordered letters (e.g., a, b, c) may be used to indicate the course sequence in the week.
Note that the specific styling of the file names (e.g.~all lower case, separated by hyphens) adheres to the tidyverse style guide which we will discuss further in Section \ref{style-guide}.
Last, but not least, the consistency of names across different file types is important.
For instance the quiz and lecture files in Figure \ref{fig:directory} are consistently named.
It is evident from the file names that \texttt{lec-11-simple-linreg.Rmd} and \texttt{quiz-11-a-simple-linreg.Rmd} go together.

\hypertarget{version-control}{%
\subsubsection{Version control}\label{version-control}}

A version control system records changes to a file or set of files over time so that changes can be tracked and specific versions of a file can be recalled later.
Version control is an important foundation for reproducible workflows, be they collaborative (working with others) or non-collaborative (tracking file histories).

Regardless of whether version control is part of the course curriculum or not, we strongly recommend adopting the use of version control into one's workflow of preparing and organizing teaching materials.
This is especially true if you're already using literate programming as suggested in Section \ref{literate-programming}.

\begin{figure}
\includegraphics[width=1\linewidth]{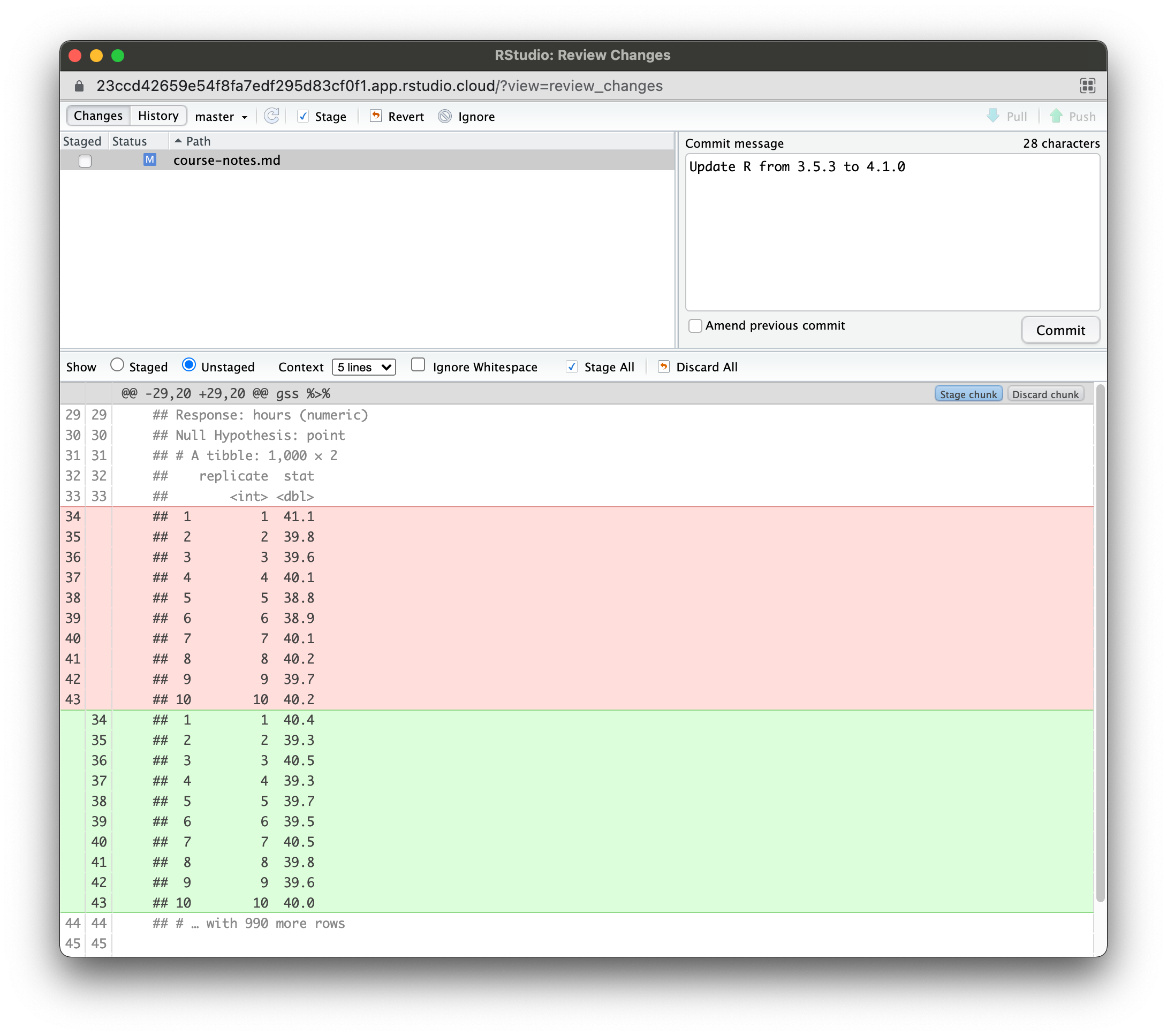} \caption{Diff due to upgrading R from 3.5.3 to 4.1.0.}\label{fig:random-number-diff}
\end{figure}

There are a few reasons for this recommendation.
First, if you use literate programming to generate course materials (e.g., slides, course notes, etc.) using a tool like R Markdown, next time you teach the course you can knit the document and use the \emph{diff} (the difference between the current state of the document and the previous state in git history, represented in a visual way highlighting changes) to see if any function output has changed with updates to packages that may have taken place in between the two semesters of teaching the course.
Looking for changes visually is difficult and cumbersome, this is an automated way to see the changes and adjust course materials accordingly.
Figure \ref{fig:random-number-diff} shows an example of such a change in output that would be difficult to detect without this explicit diff.
This is the result of a 1,000 bootstrap means taken from a sample, previously computed with R 3.5.3 and then computed with 4.1.0.
Even though a seed was set in the computation, the random number generation cannot be reproduced due to a change in the random seed generation in R 3.6.
Similarly, diff also allows instructors to detect any changes to warnings, messages, default argument selections of R functions.

Second, if you're co-teaching a course, using a web hosting service like GitHub, which pairs nicely with the use of version control with Git, is a great collaboration platform.
Co-instructors can easily see changes made by the other one, or propose changes by a formal \emph{pull request} mechanism.

We recommend hosting source files for your teaching materials in public repositories for other instructors to have access to -- in the name of furthering open educational resources.
Development of such resources doesn't have to mean writing an open textbook, which can be a huge undertaking.
Simply sharing the source code of your slides, homework assignments, etc. can go a long way in others benefiting from your teaching innovation, and they will be more likely to contribute back with feedback.
This is one way of having some impact in your academic community, which can be important for valuation purposes as well.
Additionally, constructive feedback will help better your own teaching resources.

However, we generally do not recommend sending students to the source code of the teaching materials as the first stop, particularly in introductory courses where the source code to generate output with a nice layout might be too complex for students who are new to programming to work their way through.
Share the resulting output with your students on a website, which you can easily host on GitHub too.
But you can also mention to the students that the source code is there.
Providing extra credit for spotting typos and opening issues / submitting pull requests can introduce the students wanting to take that step to workflows that is good for them to know about (even if the particular course you're teaching does not focus on version control).
It's also helpful to you.

Web hosting services like GitHub also offer project management tools that can be helpful in the context of teaching.
For example, opening issues in your own course repository is perhaps the most convenient way of quickly recording an issue you spot in your materials (e.g., you're in the classroom teaching, you spot a mistake, open an issue to record it).
Unlike sticky notes that quickly disappear, an issue will stay open until you explicitly close it.
If you don't close it that semester, it'll be a nice reminder when you pick up the same course material to teach in a subsequent semester.
Or if you're handing your materials over to a colleague who will be teaching that course, they will be able to see any notes you've left for yourself.

While version control with Git and web hosting, publicly, on GitHub is our recommendation, there are a few words of caution we should mention.
Course instructors generally keep private student files (e.g., student personal identifiers, grades, etc.) in the same folder as their teaching materials.
If you're using version control on the entire course folder, use \texttt{.gitignore} to omit from version control grading related files.
Alternatively, you put these in a separate folder that is version controlled locally, but not pushed to a public repository.
Doing this as you set up your course folder is important.
It's a bad, but a common, habit to \texttt{git\ commit} all files with changes in them and then push them to the repository, and once a file has been committed, removing it from the version control history is not trivial.
It's better to silo private student information first, before you start commits on your course folder/repository.

Another consideration might be file size limits on GitHub.
GitHub has a file limit of 100MB, and if you attempt to add or update a file that is larger than 50 MB, you will receive a warning from Git.
If you're teaching with datasets larger than this size, hosting your course materials on GitHub can be requires additional consideration.
GitHub offers large file storage (Git LFS) for free as part of their education benefits and educators can take advantage of this benefit by getting their academic employment verified.
Once verified their GitHub organization as a GitHub classroom organization, instructors who want to use LFS can use instructions on GitHub Education Community forum \citeyearpar{githubforum}.

\hypertarget{documentation-1}{%
\subsection{Documentation}\label{documentation-1}}

\hypertarget{data-documentation}{%
\subsubsection{Data documentation}\label{data-documentation}}

Data files used in teaching materials should be documented to reflect both the contents of the datasets and their provenance.
A simple and recommended way of storing this information is a plain text README file that includes both the provenance and the variable description information.
Another approach is a CSV (comma separated values) file for the variable descriptions, organized as a spreadsheet, and an additional plain text README file containing data provenance information.
Importantly, these need to be plain text files that can be version controlled, so we discourage the use of spreadsheets like an Excel file or a Google Sheet.

\hypertarget{folder-documentation}{%
\subsubsection{Folder documentation}\label{folder-documentation}}

The course folder as well as each of its top level sub-folders should have a README file that explains what is included in that folder.
Additionally, these README files should outline any steps necessary to reproduce the materials included in their respective folders, e.g., knit the document using \texttt{rmarkdown::render()}, render the webpage using \texttt{blogdown::build\_site()}.

\hypertarget{software-documentation}{%
\subsubsection{Software documentation}\label{software-documentation}}

A simple solution for software documentation is to include the list of packages being used and their version numbers as well as the version of R at the time of developing the teaching materials.
This can be done in the README file at the root of the project folder.
If the README file is being generated with R Markdown, one can use the \texttt{session\_info()} function to compile this information.
This function will print out a nicely formatted message that includes the version of R and operating system as well as each package, their versions, the version date, and the source (CRAN or repository of development version fo package).
One might also consider including this information at the end of each individual teaching document (e.g., slide deck, homework assignment) if doing so at the project level doesn't provide sufficient level of detail for reproducibility for all of the projects' contents.

On the other end of the computational reproducibility spectrum, one might consider creating a snapshot of the computational environment they are using to generate their teaching materials using Docker containers \citep{docker}.
However this solution is not trivial to implement and they can also be more difficult to maintain.

A more useful place to spend energy on is using tools that are focused on preserving the specific computational environment of the language being used, e.g., the \textbf{renv} package for R \citep{renv} and the \textbf{venv} \citep{venv} package for creating virtual environments for Python.
These solutions are particularly good about documenting the specific versions of the programming language and all packages being used in the project folder and do so in a way that is portable, i.e., they provide tooling for someone else to reproduce your work using the same computational environment of the language.

So far we have discussed documenting software that is used to generate teaching materials.
However, as educators, our communication with students does not happen only through these materials.
Nowadays just about every course has an online discussion forum where students ask questions.
In statistics and data science courses many of these questions are related to syntax errors and coding challenges that students encounter.
We recommend instructors stick to the reproducibility principle when answering these questions.
Using the \textbf{reprex} package instructors can generate reproducible code snippets and even share their session information along with the code snippet easily \citep{reprex}.

\hypertarget{style-guide}{%
\subsubsection{Style guide}\label{style-guide}}

Consider the following line of code for fitting a linear regression model:

\begin{Shaded}
\begin{Highlighting}[]
\NormalTok{modelweight}\OtherTok{\textless{}{-}}\FunctionTok{lm}\NormalTok{(mpg}\SpecialCharTok{\textasciitilde{}}\NormalTok{wt,}\AttributeTok{data=}\FunctionTok{subset}\NormalTok{(mtcars,am}\SpecialCharTok{==}\DecValTok{1}\NormalTok{),}\AttributeTok{na.action=}\NormalTok{na.exclude)}
\end{Highlighting}
\end{Shaded}

Now consider the next code block that is only different in terms of style and nothing else.

\begin{Shaded}
\begin{Highlighting}[]
\NormalTok{model\_weight }\OtherTok{\textless{}{-}} \FunctionTok{lm}\NormalTok{(mpg }\SpecialCharTok{\textasciitilde{}}\NormalTok{ wt,}
  \AttributeTok{data =} \FunctionTok{subset}\NormalTok{(mtcars, am }\SpecialCharTok{==} \DecValTok{1}\NormalTok{),}
  \AttributeTok{na.action =}\NormalTok{ na.exclude}
\NormalTok{)}
\end{Highlighting}
\end{Shaded}

The latter code is easier to read as it employs plenty of spaces and spans over multiple lines giving the reader an opportunity to focus on one aspect of the code at a time.
This allows for code to be human-readable in addition to being machine-readable.
The latter code is written based on the tidyverse style guide \citep{tidystyle}. Since we teach our courses using the packages from the tidyverse, the tidyverse style guide is a natural fit for us. The tidyverse style guide is extensive and includes formatting standards including naming objects, spaces, spanning code over multiple lines. Some other examples of style guides include Google's Python Style Guide \citep{googlestyle} and the Julia Style Guide \citep{juliastyle}.
We recommend that instructors pick an existing styles for themselves rather than developing their own set of guidelines so that what the students are exposed to in class in terms of coding style matches what they will outside of the classroom setting.

\hypertarget{openness-1}{%
\subsection{Openness}\label{openness-1}}

Releasing teaching materials openly should be the standard however many practices promoted by universities make this difficult.
For example, universities generally recommend posting teaching materials on their course management system that requires university authentication to access.
There is a good reason for this suggestion -- students like having some consistency in how they access course materials across the various courses they take.
Even if you choose to post your materials in such venues to share with your current students, you might consider also posting them publicly for the rest of the world to see (open access) or for other educators to adapt and reuse (open source).

\hypertarget{licensing}{%
\subsubsection{Licensing}\label{licensing}}

Regardless of your level of openness (open access to open source), you should release your materials with a license.
The particular license you choose for your teaching materials depends on the type of materials you're distributing (lesson plans and slide decks vs.~code) as well as your personal choices for how you would like the material to be reused by others.

For teaching materials that are not software we recommend using a Creative Commons license \citeyearpar{CClicense}. The Attribution-NonCommercial 4.0 International (CC BY-NC 4.0) is commonly preferred by educators who want to limit commercial use of their materials and require attribution when reused. This license is particularly useful if you might publish your materials as a textbook as it prevents others from reusing your work in a commercial setting (e.g., publication by a for-profit publisher). If you would like to allow usage of your materials in commercial settings as well (e.g., corporate training, massive open online courses offered by for-profit companies), you might prefer Attribution 4.0 International (CC BY 4.0). Another consideration in licensing is how (if at all) derivatives can be generated and re-shared (e.g., with a share-alike option). It's important for educators to inform themselves on the various options and further details of these options before choosing a license. For software-based teaching materials (e.g., an R package for teaching), the MIT license \citeyearpar{MITlicense} and the General Public License \citep{GPLlicense} are more appropriate than Creative Commons licenses.
For readers interested in understanding the relationship between licensing and open educational resources further we recommend the manuscript by \citet{hilton2010}.

\hypertarget{hosting}{%
\subsubsection{Hosting}\label{hosting}}

Sharing teaching materials publicly is best done via public course websites.
These websites need to be hosted and deployed for the students and the general public to have access.
There are many options for hosting course websites.
For instance, GitHub pages is a feature that allows a website hosting directly from a GitHub repo.
Similarly, Netlify allows hosting statistic websites and can handle continuous deployment from a GitHub repository.
In other words, any changes made to a GitHub repo can be reflected on a Netlify website in a matter of a short time.
We have found free tiers of Netlify and GitHub pages to be sufficient for deploying our course websites.
Last but not least, hosting course websites on institutional websites is an option.
Although, the server maintenance schedule of institutional websites may interfere with website hosting.

\hypertarget{examples}{%
\section{Examples}\label{examples}}

In this section, we share examples from our own courses that are open-access and open-source.
We hope the examples we provide here can provide a starting point for instructors who are new to reproducible teaching.
Both of us teach introductory data science courses and our websites are publicly available at \mdids{} and \mcrids{}.
In both these courses, we follow the framework presented here thus detail the examples closely following the framework.

In addition to the website interface that students often interact with, we also share the source code for all materials (e.g.~slides, R code etc.) posted on these websites, along with the source code for making the websites, hosted on GitHub and linked from these pages as well.
The GitHub repos available at \mdidsgithub{} and \mcridsgithub{} respectively can be good starting point for readers to see ``behind the scenes'', in other words the source code of the teaching materials.
In the first course, all course materials are organized in a single GitHub repository while in the second course the source codes for the website, slides, homework assignments, etc. are organized in separate repositories that all live in a single GitHub organization.
Depending on the amount and complexity of the course materials as well as personal preference for file organization, instructors can choose either approach for organization as long as there is no ambiguity about where the source code for certain materials can be found.

In terms of computational reproducibility, we make use of R Markdown and its extension packages to develop our course materials.
For instance, one can notice the slides folders and their sub-directories include several files with \texttt{.Rmd} extension representing R Markdown files.
When we use external data, we keep the raw data as is (often including the file name).
An example can be in the \texttt{slides/data} folder of the first course.
Overall in both courses' repos, file and folder names have clear enumeration as well as words that represent the content.
Last but not least, these repos are hosted on GitHub which reveals the git version history.
Readers can access the history by adding \texttt{/commits/main} to the URLs of the repos.

In terms of documentation, one may immediately notice the \texttt{README.md} files for each repo. GitHub by default displays \texttt{README.md} files when a repo link is visited.
For example, the README of the course website for the second course provides an orientation to the course materials, clearly outlines where the source code for each component can be found, and also provides guidance for reuse of course materials.
We also make use of README files for data documentation (e.g., data provenance and dictionary information) and save these in folders where datasets appear in the course materials.
All file and folder names follow the aforementioned the Tidyverse style guide using all lower case file names separated by hyphens.
Within the \texttt{.Rmd} files, the R code used and presented to students also follow this style.

Needless to say, both the course websites are open-access and open-source.
All course repos feature \texttt{LICENSE.md} files which detail the terms of use for these materials.

In addition to the two aforementioned courses, more broadly \dsbox{} is an example of an open access, open source repository that is designed specifically for other educators to adopt and reuse the materials.
We encourage interested readers to check it out and notice the similarities in terms of the reproducible teaching framework.

\hypertarget{discussion}{%
\section{Discussion}\label{discussion}}

The framework we have presented here can be considered as a checklist for attaining reproducible practices in teaching.
We are aware that the list is extensive.
Similar to any new approach in teaching, the reproducible teaching framework also comes with its own learning curve and can initially be overwhelming for anyone who is adopting to the tools of this framework for the first time.
Thus, we recommend instructors to make incremental changes from one academic term to the next and adopt reproducible practices gradually.

A major component of the toolkit we present here has been developed in the last decade.
We had to self-teach how to use these tools and adopt them for teaching purposes.
For the new generation of statistics and data science educators, adoption of these tools in their training programs is extremely crucial and the benefits of teaching-focused training programs (e.g., workshops, courses) can extend to their work beyond teaching \citep{rummerfield2021}.
We recommend the technical aspects of teaching training in statistics and data science to include exposure to reproducible teaching toolkit(s).

Without a doubt, reproducibility practices in teaching have high impact on course management.
For instance, in one of our courses, we have used Git for version control, Gradescope for grading assignments, Canvas as the learning management system to store grades, RStudio Cloud for coding interface, course website for dissemination of information, GitHub for hosting the course website, and Piazza for discussions.
Without organized conventions such as file naming and folder structures, managing courses with multiple tools, hundreds of students, and their assignments would have been chaotic.
Reproducibility practices not only help the instructors reproduce their or someone else's teaching materials, but they also help instructors navigate across different tools while having learners also navigate easily.

The impact of teaching can be extended beyond the classroom when open practices are adopted.
When teaching materials are shared publicly they not only benefit other instructors who may be teaching similar courses, they they also provide free access to learning materials to learners from around the world.
In addition, sharing teaching materials openly can also benefit the instructor by helping them gain name recognition in their field as well as leading to potential collaboration opportunities, which can be especially impactful if they are early career \citep{dogucu2021}.
Instructors who may want to disseminate their teaching materials can consider using social media accounts, American Mathematical Society's Open Math Notes \citeyearpar{ams}, Learn R for Free website \citeyearpar{learnR4free}, and RStudio Education's GitHub repo for courses teaching R \citeyearpar{rstudioed}.

We envision modern statistics and data science instructors to adopt the reproducible teaching framework in their own teaching regardless of the nature of the courses that they are teaching.
However, we also acknowledge that the benefits of adopting this framework can differ based on the course being taught.
For example, consider a data science course or a statistical consulting course where students work on projects and use reproducible workflows for their work.
In such courses, the consistency between the instructors' and students' practices would be highly beneficial as the instructor sets a role model for the students.
In courses, where the focus is not necessarily reproducible workflows (e.g.~a mathematical probability theory course) the instructors can still adopt a reproducible workflow for their own teaching.
The benefits of this workflow for the instructor, other members of the teaching team, as well as the broader data science education community are still highly valuable, despite not serving as a role model for the students.

\bibliographystyle{agsm}
\bibliography{bibliography.bib}

\end{document}